\documentclass{aa}
\usepackage{graphicx}

\begin{document}

\title{Far-infrared mapping of the starburst galaxy \object{NGC 253} with ISOPHOT.
\thanks{Based on observations with ISO, an ESA project with instruments funded 
by ESA Member States (especially the Principal Investigator countries: France,
Germany, Netherlands and the United Kingdom) and with the participation of 
ISAS and NASA.}
}

\subtitle{}

\author
{M. Radovich\inst{1}, J. Kahanp\"a\"a\inst{2} \and D. Lemke\inst{3}}
\institute
{Osservatorio Astronomico di Capodimonte, 
via Moiariello 16, I-80131 Napoli, Italy\\
\email{radovich@na.astro.it}
\and 
  Observatory, University of Helsinki, PL 14, 00014 Helsingin yliopisto, 
Finland\\
\email{jere.kahanpaa@helsinki.fi}
\and 
  Max-Planck-Institut f\"ur Astronomie, K\"onigstuhl 17, D-69117 Heidelberg,
Germany\\
\email{lemke@mpia-hd.mpg.de}
}
\authorrunning{M. Radovich et al.}
\titlerunning{Far-infrared mapping of \object{NGC 253}}
\offprints{M. Radovich}

\date{Received xxx /Accepted xxx}

\abstract{
A 180\,$\mu$m map and strip maps at 120 and 180\,$\mu$m  were obtained  for
the edge-on starburst galaxy \object{NGC 253} with ISOPHOT, the  photometer on
board the {\em Infrared Space Observatory}. We compare these observations with
those obtained by IRAS at 60\,$\mu$m and 100\,$\mu$m and derive the
far--infrared spectral energy distribution at different locations in the
galaxy. There is evidence for the presence of cold dust (T $\le$ 20 K) both in
the nucleus and in the disk. Extended emission dominated by cold dust is
detected up to $\sim $ 15\arcmin\ ($\sim 10$ kpc) 
along the major and minor axis; its spatial distribution is similar to that seen 
in the IRAS and ROSAT PSPC images. The emission along the minor axis
is  probably related to large-scale outflows of gas (superwinds) which
originate  in the nuclear starburst and maybe to star formation in the halo. 
The radial dependence of the dust temperature 
along the major axis is found using a radiative transfer code: we show that the dust 
scale length in the disk is $\sim$ 40\% larger than that of stars.
}

\maketitle

\section{Introduction}

There is growing evidence (Trewhella et al. \cite{trewhella}) for the
existence of two components in the dust distribution in galaxies, a spatially
extended distribution of cold dust (T $<$ 20K) and a warm dust (T $>$ 20K) 
component concentrated in the disk, probably associated with star formation. 
The most direct evidence for the existence of the cold dust component has been 
provided by SCUBA and  
by ISOPHOT, the photometer on board the {\em Infrared Space Observatory}
(ISO, Kessler et al. \cite{kessler}), in particular thanks to the C200 
detector.
The cold dust may lie both along the disk, in which case the dust scale length 
may be higher than the stellar scale length by factors of 40\% to 80\% 
(Alton et al. \cite{alton98}b) and vertically.  If a massive nuclear starburst
is present, supernovae forming there  may drive the gas from the 
inner disk of the host galaxies to their halos, as first proposed by
Chevalier \& Clegg (\cite{chevalier}) and later Heckman et al. (\cite{ham}). 
Dust entrained in these outflows should enhance the far-infrared emission
vertically to the disk; this effect should be most easily  observed in 
edge-on galaxies. 

\object{NGC 253} is a nearby, almost edge--on barred spiral galaxy 
(De Vaucouleurs et al. \cite{devauc}); we adopt a distance of 2.5 Mpc
(Houghton et al. \cite{houghton}), which gives a projected linear scale
of 12\,pc/arcsec. \object{NGC 253} is one of the nearest galaxies showing evidence
for a compact nuclear starburst (size $\sim$ 100 pc, Dudley \& Wynn-Williams
\cite{dudley}) which is partially responsible for its high far--infrared
luminosity, L$_{\rm FIR} \sim 2\times10^{10} L_\odot$. 
\object{NGC 253} has been extensively observed in all spectral ranges (see e.g. 
Engelbracht et al. \cite{engelbracht} and Forbes et al. \cite{forbes}). 
Still uncertain is the presence of a highly obscured AGN which would contribute 
some of the radio and X-ray emission, but none of the optical and infrared
emission (Forbes et al. \cite{forbes}).

McCarthy et al. (\cite{mccarthy}) found along the minor axis an extended 
emission-line region associated with diffuse x--ray  emission, over 
$\sim$ 15\arcsec\ in diameter: they identified this emission-line gas with 
cool dense clouds embedded in an outflowing wind of hot gas. 
Fabbiano (\cite{fabbiano}) found more extended, 
low surface brightness X-ray emission extending along the minor axis out to 
10\arcmin\ (NW) and 5\arcmin\ (SE). More recently, Dahlem et al. 
(\cite{dahlem}) analyzed ROSAT PSPC and HRI and ASCA data and found a diffuse, 
hot soft X-ray halo with dimensions of $16\times10$ kpc and attributed it
to  the superwind. Pietsch et al. (\cite{pietsch}) also investigated the
extended X-ray emission from ROSAT HRI and PSPC data: they separated the
total halo emission in a corona component which originates from the halo
immediately above the disk (scale height $\sim$ 1 kpc) and in an outer halo
which extends to projected distances from the disk of 9 kpc and shows a horn-like
structure. 
However, IRAS images of \object{NGC 253} did not provide straightforward results. Rice
(\cite{rice}) and Alton et al. (\cite{alton-farir}a) presented high--resolution
IRAS maps produced with the Maximum Correlation Method (MCM) technique; they
concluded that spurious effects like detector hysteresis at 12 and 25\,$\mu$m
and emission reflected from the IRAS telescope spider at 60 and 100\,$\mu$m
mask low--level emission outside the disk. 
Alton et al. (\cite{alton-submm}) presented submillimeter images of the nuclear 
regions of \object{NGC 253} obtained with SCUBA and found evidence for a dust 
outflow along the minor axis at 450\,$\mu$m on a size $\le 45\arcsec$.

In this paper we report the results of observations of \object{NGC 253} obtained 
with ISOPHOT; available data and their processing are discussed in
sec.~\ref{sec-obsred}. The investigation aims at: 
\begin{enumerate}
\item Evaluating the far-infrared emission properties in different regions of 
the galaxy and look for evidence of the outflow in the FIR. To this purpose 
we combine the ISOPHOT 180\,$\mu$m map with the IRAS maps 
(sec.~\ref{sec-results}): an estimate of dust temperatures and masses is 
obtained by fitting the so-obtained spectral energy distributions with a simple 
model (sum of two modified blackbodies).  
\item Analyzing the radial dependence of the dust temperature using a radiative 
transfer code (sec.~\ref{sec-models}). FIR emission properties in the disk are
compared to those derived from optical data, 
which are mainly related to the stellar component.
\end{enumerate} 

\begin{figure}
\resizebox{\hsize}{!}{\includegraphics[bb=70 100 535 570,angle=270,clip]{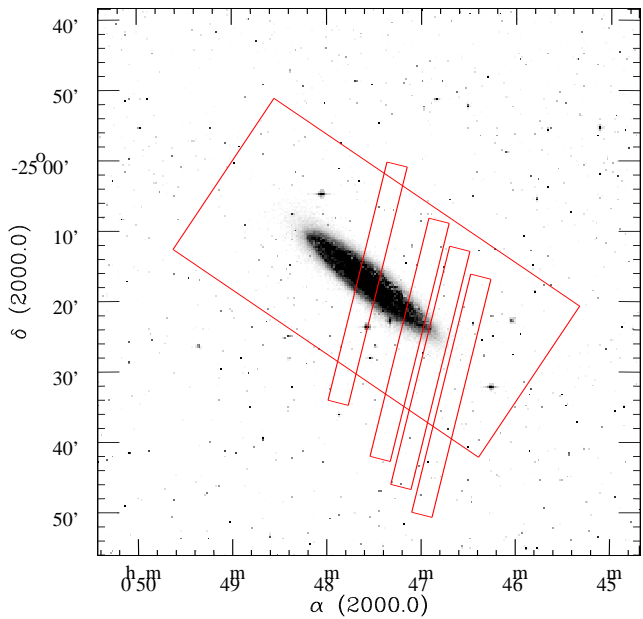}}
\caption{The areas covered by ISOPHOT maps have been overlaid 
 to J band image 
of NGC 253 from the Digitized Sky Survey, obtained using the SkyView 
survey analysis system.} 
\label{fig:isodss}
\end{figure}

\begin{table*}
\caption{Log of the ISOPHOT observations. The position angle (PA) is from 
North to East. Offsets from the nucleus are given for the strip maps.} \label{tab:log}
\begin{flushleft} 
\begin{tabular}{lc ccc cc ccc}
\hline 
$\lambda$ ($\mu$m) & $\frac{\lambda_c}{\Delta\lambda_c}$ & offset & M &  N & $\delta M$
& $\delta N$& fov & PA & exp (s)\\ \hline
120 & 2.4 & 0 &  11 &1 & 180\arcsec &  0 & 35\arcmin x 3\arcmin &  166\degr &
960\\
120 & 2.4 & 10\arcmin &  11 &1 & 180\arcsec &  0 & 35\arcmin x 3\arcmin &
166\degr & 960\\
120 & 2.4 & 15\arcmin &  11 &1 & 180\arcsec &  0 & 35\arcmin x 3\arcmin &
166\degr & 960\\
120 & 2.4 & 20\arcmin &  11 &1 & 180\arcsec &  0 & 35\arcmin x 3\arcmin &
166\degr & 960\\
 180 & 2.6 & 0\arcmin &  11 & 1 & 180\arcsec & 0 &  35\arcmin x 3\arcmin &
166\degr & 960 \\
180 & 2.6 & 10\arcmin &  11 & 1 & 180\arcsec & 0 &  35\arcmin x 3\arcmin &
166\degr & 960\\
180 & 2.6 & 15\arcmin &  11 & 1 & 180\arcsec & 0 &  35\arcmin x 3\arcmin &
166\degr & 960\\
180 & 2.6 & 20\arcmin &  11 & 1 & 180\arcsec & 0 &  35\arcmin x 3\arcmin &
166\degr & 960\\
180 & 2.6 &  & 8 & 17 & 180\arcsec & 180\arcsec &
26\arcmin x 53\arcmin & 34\degr &  4200 \\ \hline
\end{tabular} 
\end {flushleft}
\end{table*}

\begin{figure*}
\begin{tabular}{rr}
\includegraphics[bb=45 50 595 580,width=6.5cm,angle=0,clip]{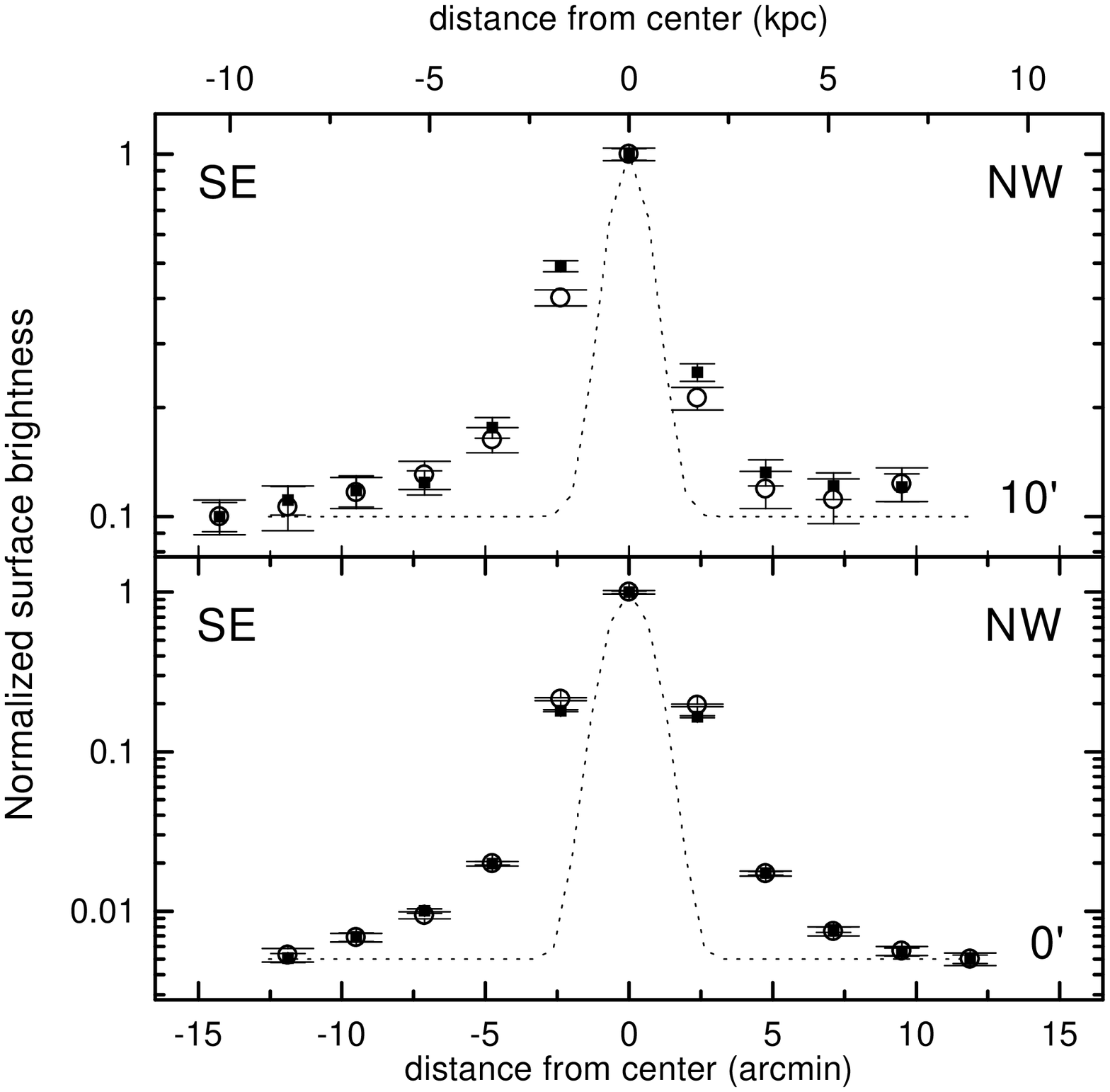}
&
\includegraphics[bb=45 50 595 580,width=6.5cm,angle=0,clip]{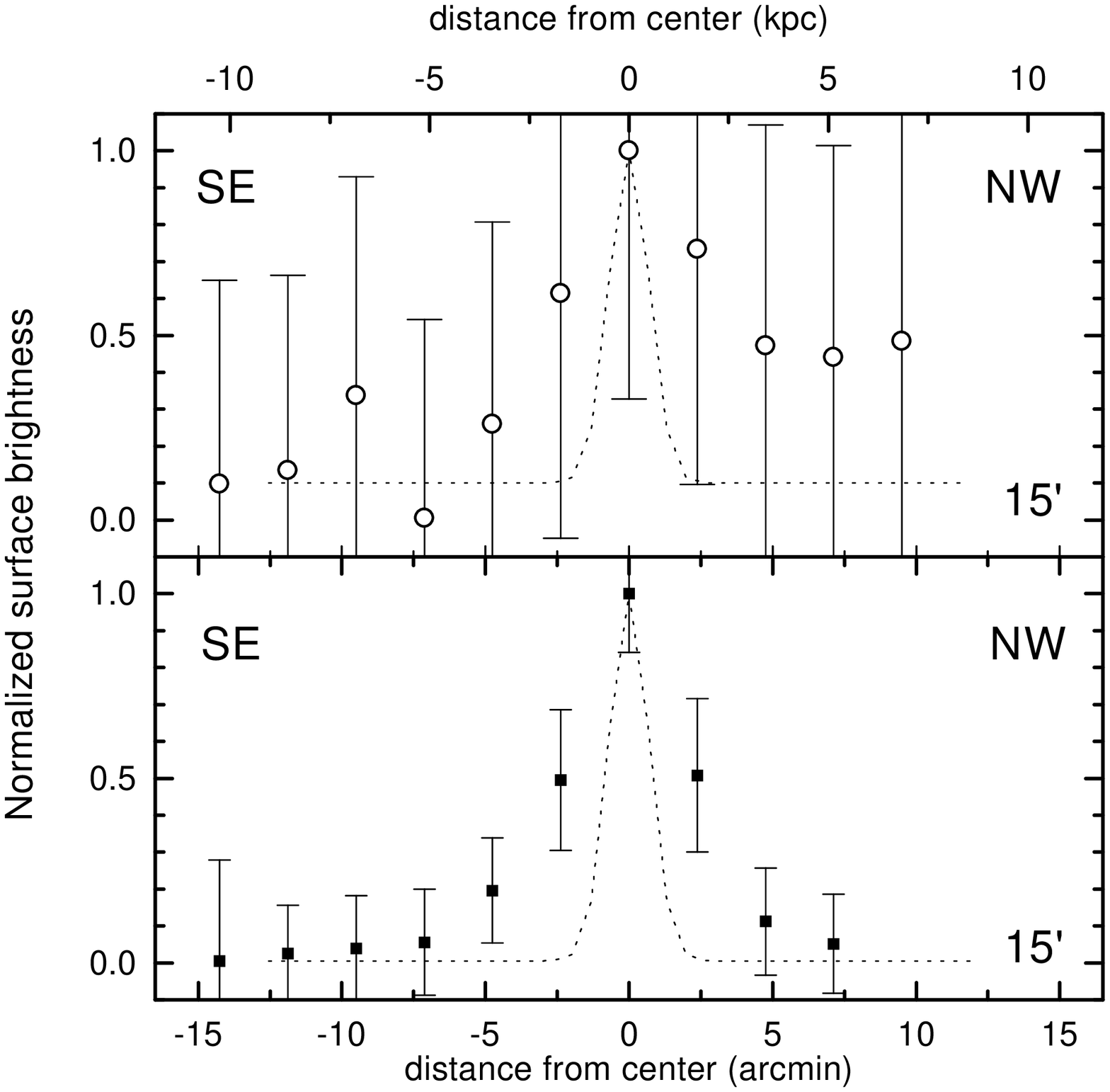}
\\
\end{tabular}
\caption{Surface brightness profiles normalized to the peak brightness in the 
strip maps at 120\,$\mu$m (open circles) and 180\,$\mu$m (filled squares) 
at offsets of 
0\arcmin, 10\arcmin\ ({\em left}) and 15\arcmin\ ({\em right}). 
The lower limit has been chosen to emphasize the extended part of the 
emission; error bars are computed from the statistical uncertainties on 
measured surface brightnesses.
The dashed line is the beam profile derived from ISOPHOT maps 
of a point--like source at 120 and 200\,$\mu$m.} 
\label{fig:sb1d}
\end{figure*}

\section{Observations and Data Reduction}
\label{sec-obsred}
\object{NGC 253} was observed with ISOPHOT (Lemke et al. \cite{lemke}) in the staring
raster mode with the multi-filter Astronomical Observation Template
PHT22 (Klaas et al. \cite{ISOman}) and the C200 detector. The staring raster
mode is a sequence of staring observations on a two--dimensional 
($M \times N$) regular grid which consists of a sequence of spacecraft pointings.
As given in Table~\ref{tab:log}, a map was obtained at 180\,$\mu$m covering
a $53\arcmin\times26\arcmin$ field.
The separation of the raster points was equal
to the C200 array size (180\arcsec), which resulted in gaps between them.  In 
addition, four strip (N=1) maps were performed in the 120 and
180\,$\mu$m filters. They were oriented roughly along the minor axis of 
\object{NGC 253} and centered on the nucleus and offset positions of 10, 
15 and 20\arcmin.
The positions of the four maps are shown in Fig.~\ref{fig:isodss}.
  
The data reduction was performed using the ISOPHOT Interactive Analysis tool 
(PIA, version 7.2) together with the calibration data set V~4.0 
(Laureijs et al. \cite{laureijs}): corrections were made for non-linearity
effects  of the electronics, disturbances by cosmic rays 
(deglitching) and signal dependence on the reset  interval time. 
The flux calibration is based on measurements with the
thermal fine calibration sources (FCS) on board: two FCS measurements were taken,
one before and one after each observation in order to correct for changes in the
detector responsivity. 
According to Klaas et al. (\cite{ISOacc}), ISOPHOT absolute calibration 
accuracies for extended sources observed in the staring raster mode are 
$\le 25\%$.
Saturation occurred in the nucleus both at 120 and 
180\,$\mu$m; this could be at least partially recovered only in the 180\,$\mu$m 
strip map including the destructive read-outs, which are usually discarded.

Our data were further processed using a series of procedures developed by Manfred
Stickel at MPIA. A first set of procedures allowed  to improve the flat field
correction of the 180\,$\mu$m map. A second set of procedures allowed to
interpolate between the detector gaps changing  the  pixel size to 50\arcsec\ 
and to rotate the map; these procedures are based on
the 'drizzle' algorithm (Fruchter \& Hook \cite{fruchter}) implemented in
IRAF.
ISOPHOT maps of a point--like source (NGC 7027) at 120 and 200\,$\mu$m were
processed following the same procedure, in order to measure the size (FWHM)
of the beam profile: in both cases we obtain FWHM = 122\arcsec.

\begin{figure*}
\begin{tabular}{rrr}
\includegraphics[bb=60 80 530 575,width=5.3cm,angle=270,clip]{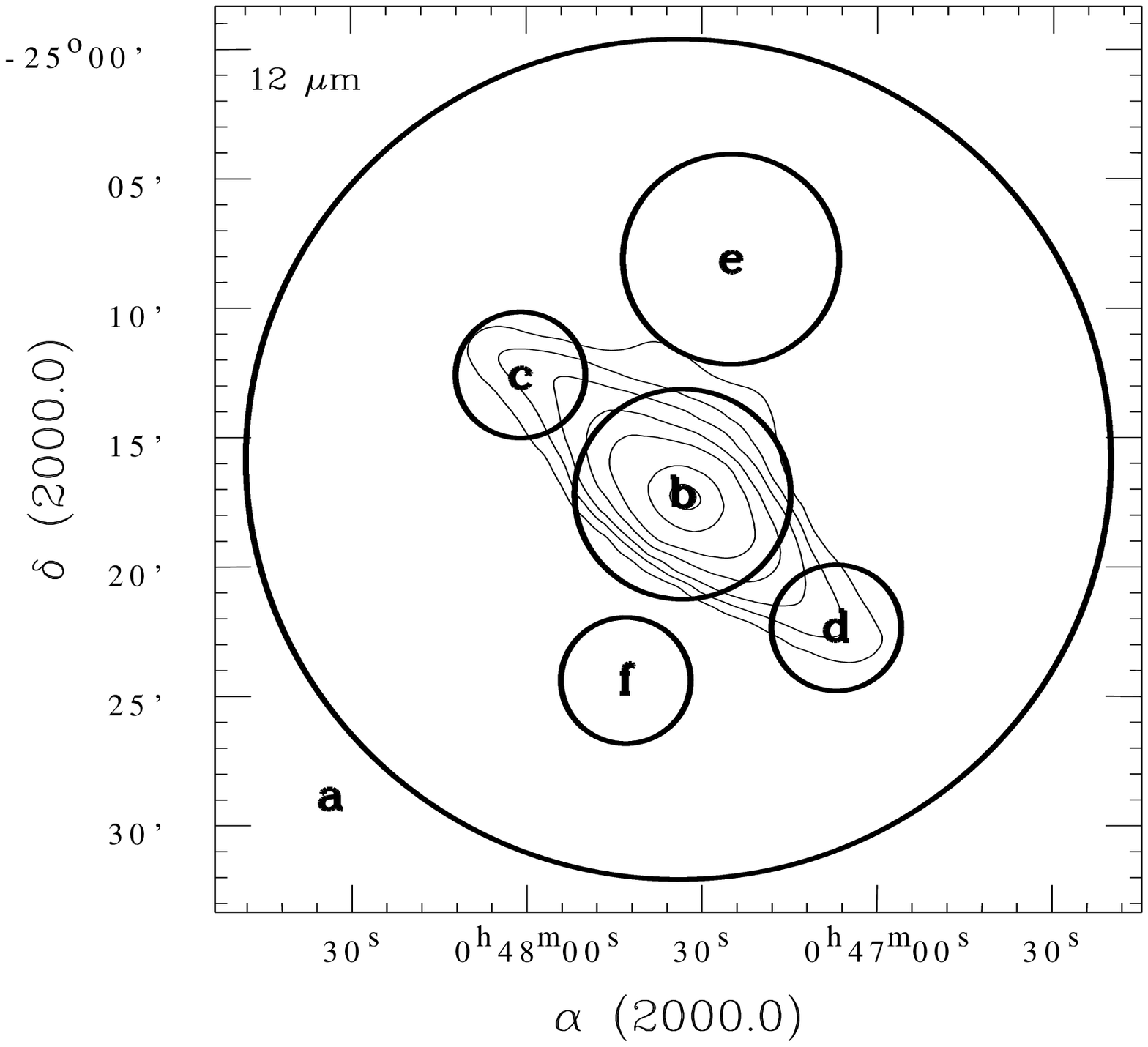}
&
\includegraphics[bb=60 80 530 575,width=5.3cm,angle=270,clip]{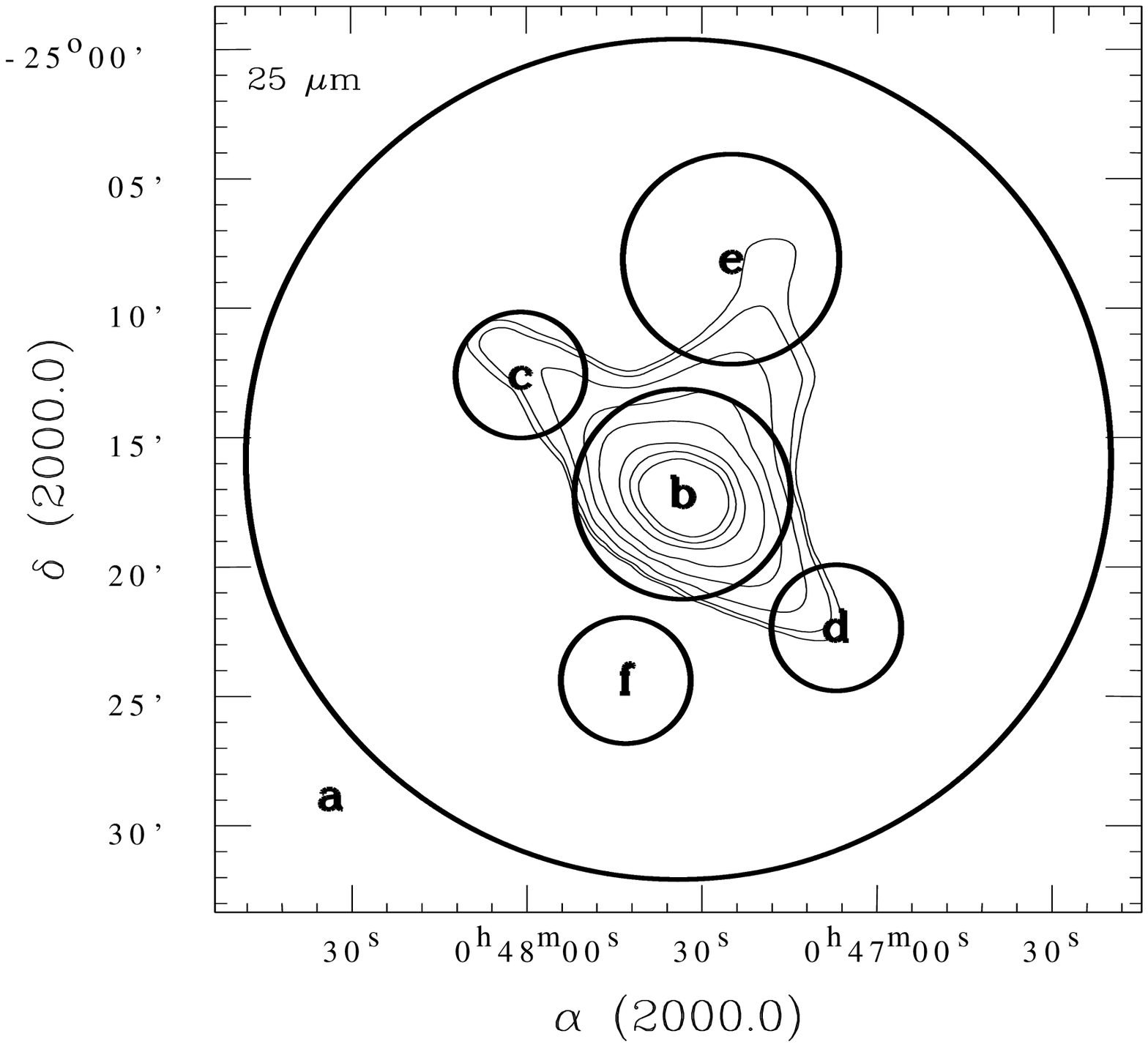}
&
\includegraphics[bb=60 80 530 575,width=5.3cm,angle=270,clip]{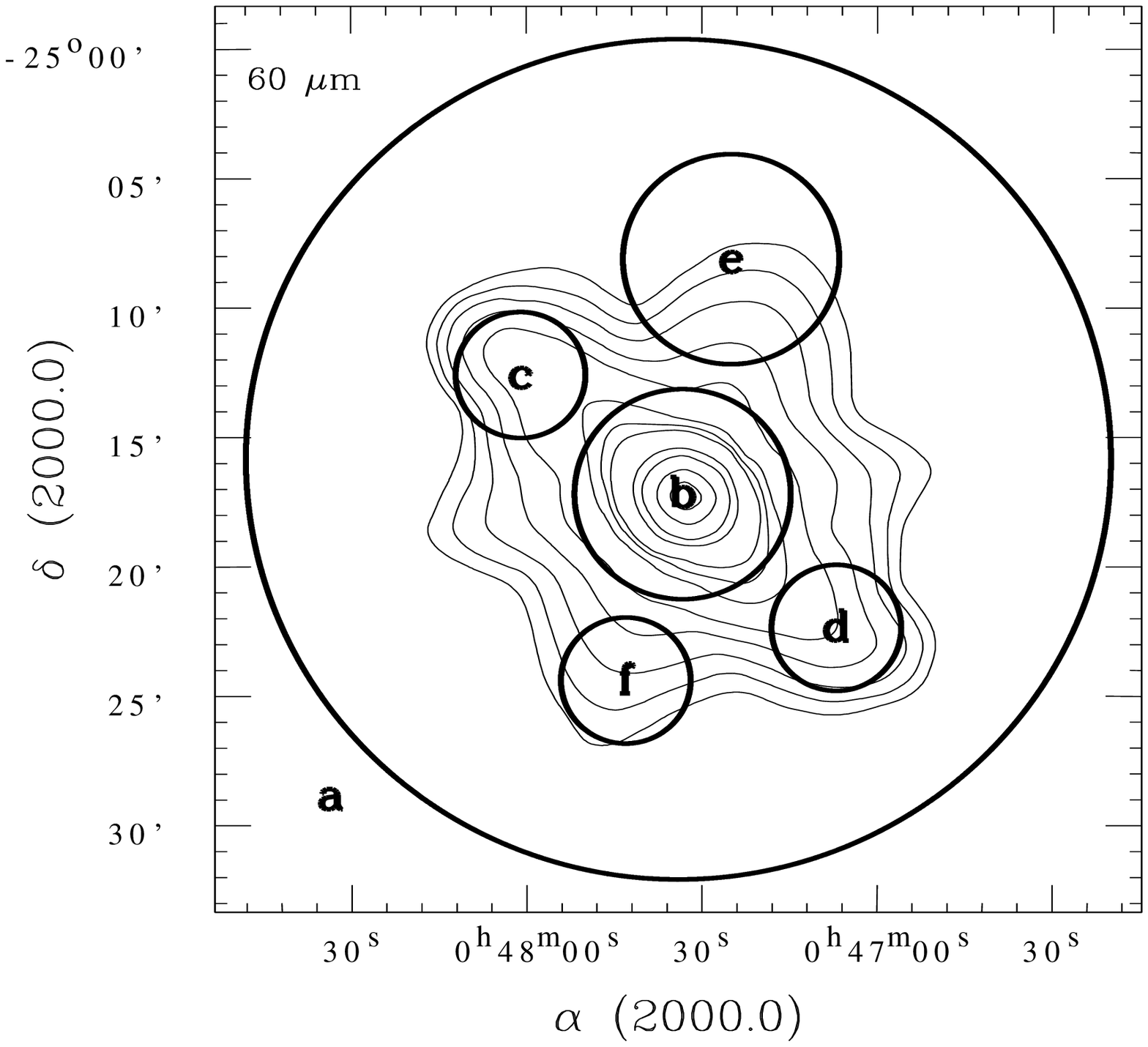}

\\
\end{tabular}
\begin{tabular}{rr}
\includegraphics[bb=60 80 530 575,width=8.1cm,angle=270,clip]{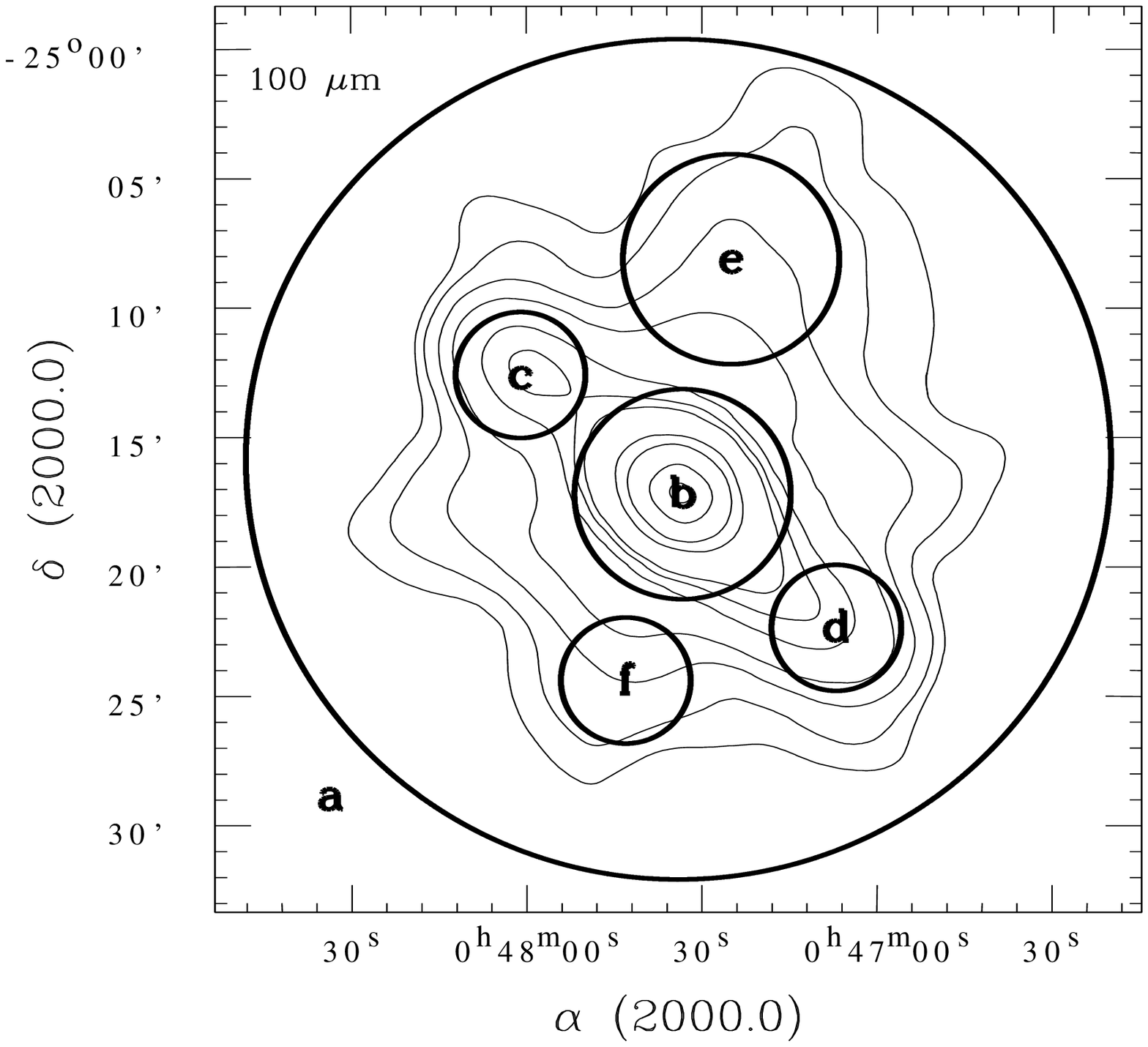}
&
\includegraphics[bb=60 80 530 575,width=8.1cm,angle=270,clip]{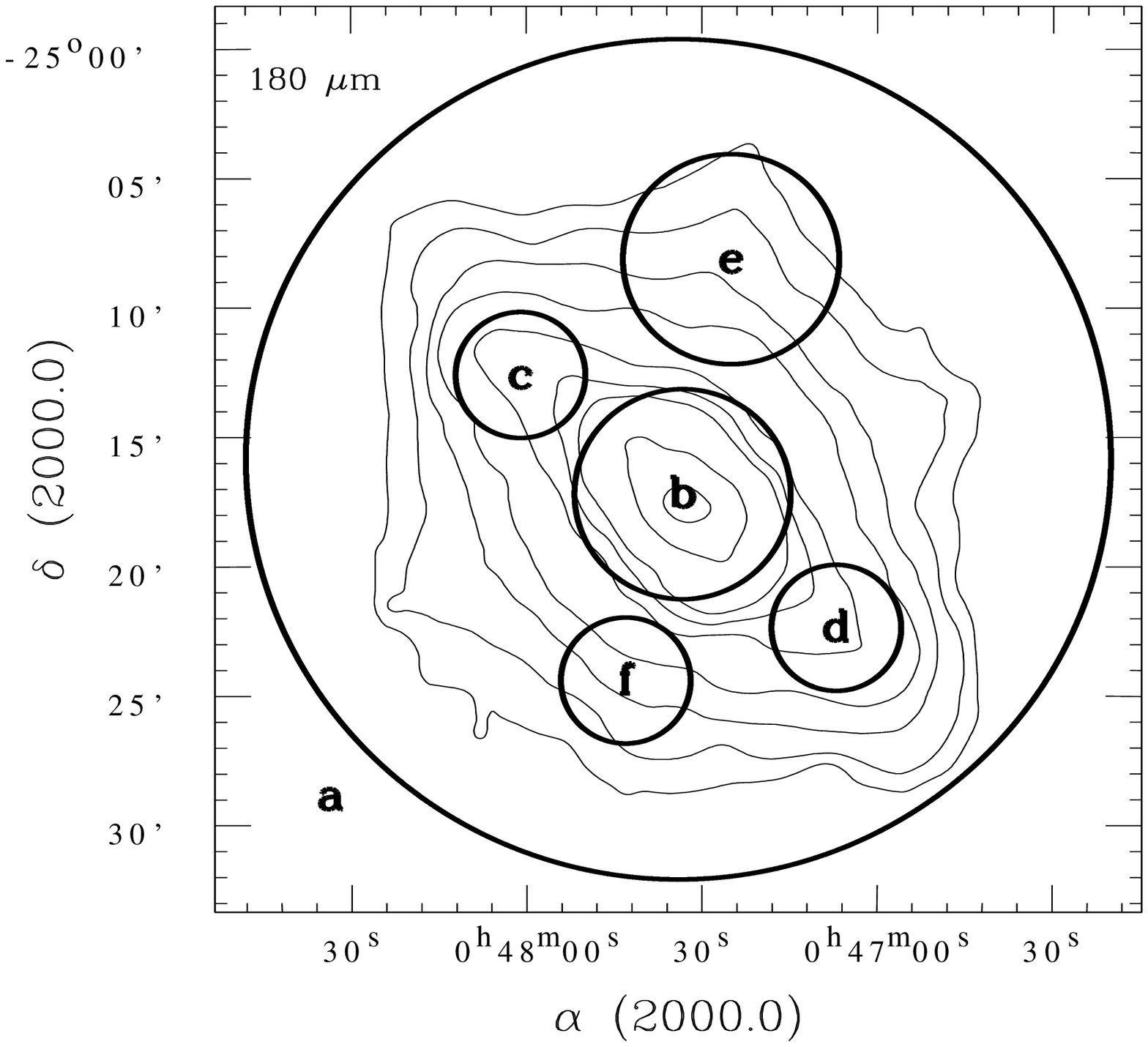}
\\
\end{tabular}
\caption{Contour maps of the IRAS and ISOPHOT images and circular apertures where 
fluxes have been measured. The IRAS maps were transformed to the same position and resolution as the ISOPHOT map.
The contour levels (MJy/sr) are: 
12\,$\mu$m - 0.8, 1.5, 2.5, 5, 10, 30, 50, 80, 300, 500, 1000, 1500; 
25\,$\mu$m - 1.2,  1.5, 2.5, 5, 10, 30, 50, 80, 300, 500, 1000, 1500;
60\,$\mu$m -  1.0,  1.5, 2.5, 5, 10, 30, 50, 80, 300, 500, 1000, 1500;
100\,$\mu$m -  1.5, 2.5, 5, 10, 30, 50, 80, 300, 500, 1000, 1500;
180\,$\mu$m - 1.5, 2.5, 5, 10, 30, 50, 80, 200, 300, 370.
}
\label{fig:dispcont}
\end{figure*}



\subsection{IRAS maps}
\label{iras}

IRAS images of \object{NGC 253} processed with the High Resolution (HIRES) technique
were obtained from the Infrared Processing and Analysis Center (IPAC). A
detailed description of this technique was given by Rice (\cite{rice}). We 
used the maps with twenty iterations, giving resolutions of 37x23\arcsec\ 
(12\,$\mu$m), 35x23\arcsec\  (25\,$\mu$m), 62x41\arcsec\ (60\,$\mu$m) and 
98x80\arcsec\  (100\,$\mu$m). 

 The IRAS maps were convolved with a Gaussian 
function to the FWHM of the ISOPHOT beam profile at 200\,$\mu$m (122\arcsec) 
and registered to the ISOPHOT 180\,$\mu$m map using the coordinate information 
contained in the headers of the image files: the registration also included 
matching the pixel size in the maps. The reliability of the 
registration was checked by the coincidence of the surface brightness peak; 
the IRAS  and ISOPHOT maps are compared in Fig.~\ref{fig:dispcont}. 

\begin{figure*}
\begin{tabular}{cc}
\includegraphics[bb=35 100 300 760,height=17cm,angle=0,clip]{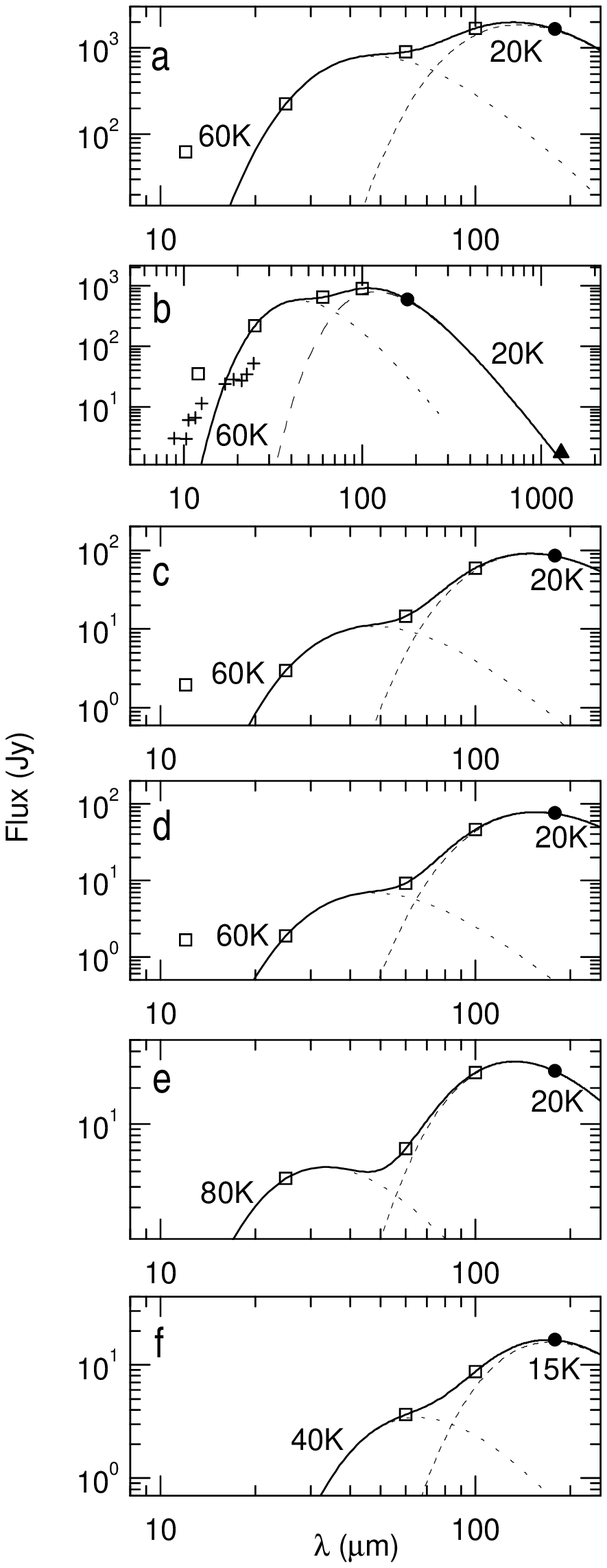} &
\includegraphics[bb=35 100 300 760,height=17cm,angle=0,clip]{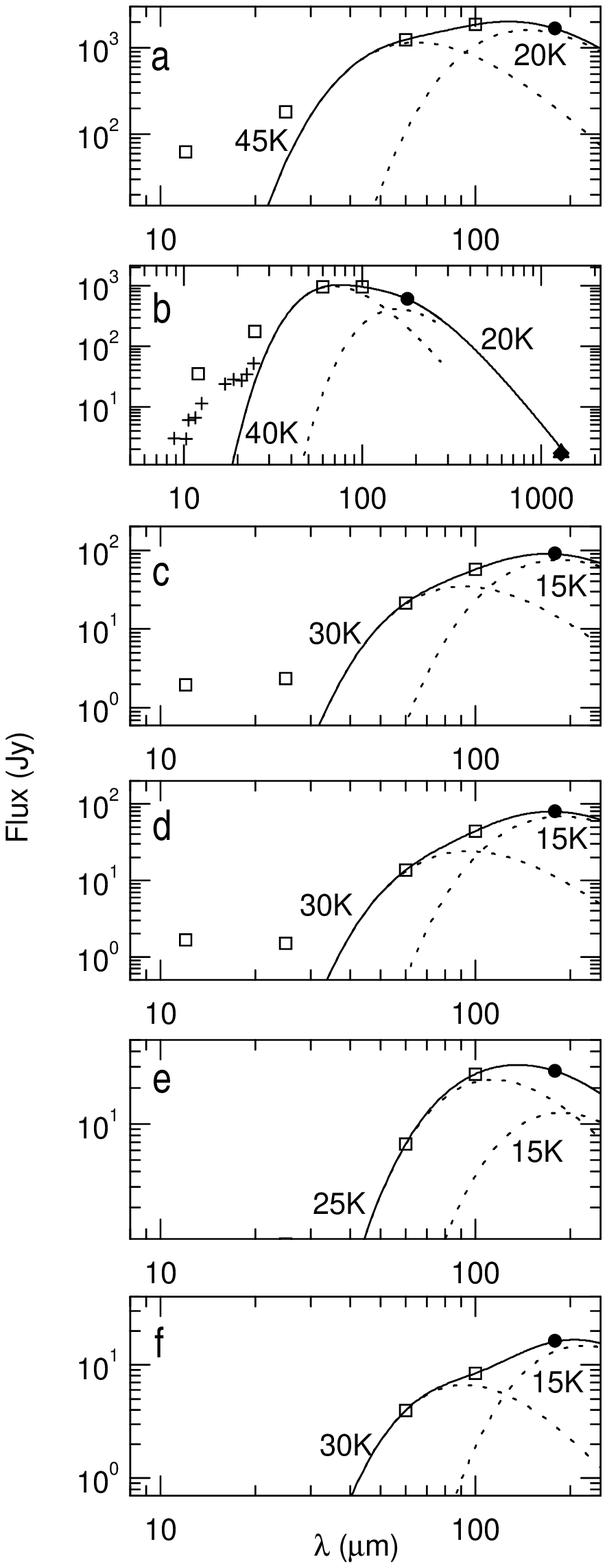}
\\
\end{tabular}
\caption{Spectral energy distributions measured in the regions displayed in Fig.~\ref{fig:dispcont};
fluxes have been color corrected. IRAS fluxes are displayed by squares,
ISOPHOT fluxes by circles; crosses show ground--based photometry of the
nucleus (5.5\arcsec\ aperture) taken from NED; the triangle is the 1300\,$\mu$m 
flux given by Kr\"ugel et al. (\cite{krugel}).}
\label{fig:seds}
\end{figure*}

\section{Mapping of the FIR emission}
\label{sec-results}
The surface brightness profiles in the strip maps show (see
Fig.~\ref{fig:sb1d}) that the emission is dominated by the strong nuclear 
source.
However, extended emission is also present along the minor axis,  
in particular in the strip centered on the nucleus (0\arcmin\ offset) 
where it may be detected within uncertainties to a distance $\sim$ 8\arcmin\ 
from the nucleus and with decreasing signal to $\sim$ 12\arcmin.
We define this emission as the
`halo' component, in addition to that detected along the major axis  
(`disk') and in the center (`nucleus').
Along the major axis, emission is detected up to 15\arcmin, where the data
at 120\,$\mu$m are significantly noisier than at 180\,$\mu$m;
no emission above the background is detected in the 20\arcmin\ strip map.
In order to
measure the fluxes in these three components, we defined circular  apertures at
different positions in the galaxy (see Fig.~\ref{fig:dispcont} and 
Table~\ref{tab:params}) and computed the integrated fluxes from the
background-subtracted  IRAS and ISOPHOT maps. 

The flux measured in the aperture centered on the nucleus is the 
sum of disk and nuclear emission. 
 Due to the poor spatial resolution it was not possible to separate the two
components in our data. We made an attempt to estimate the emission that 
would be produced by a `point-like' source (FWHM $\le$ 122\arcsec); 
we shall define this component as the `nucleus', with the 
understanding that actually it is the sum of nuclear and disk emission
inside the region that would be seen as `point-like' by the C200 detector. 
To this aim, we followed a procedure similar to that described by 
Alton et al. (\cite{alton-farir}a).
We normalized the 2D beam profiles to the 
observed peak surface brightnesses and integrated them in an aperture of 
diameter $\sim$ 8\arcmin. In the case of the IRAS images we used the beam maps 
provided by IPAC. 
In the case of ISOPHOT we used a 2D gaussian fit of the point-like
source where the beam profile was measured: this was normalized to the peak 
surface brightness found in the 
180\,$\mu$m strip map since the nucleus in the large 180\,$\mu$m map is 
saturated. The ratio of the
IRAS  total to  nuclear (rescaled beam maps) fluxes gives results similar to those found by Alton et
al. (\cite{alton-farir}a) for infrared--bright galaxies, namely
$F^{\rm tot}_{60}/F^{\rm nuc}_{60} \sim 1.4$ and $F^{\rm tot}_{100}/F^{\rm nuc}_{100} \sim 1.8$;
from the ISOPHOT map we get $F^{\rm tot}_{180}/F^{\rm nuc}_{180} \sim 3$. It follows that
while $\sim$ 70\% of the 60\,$\mu$m emission comes from the nucleus, emission at longer
wavelengths is dominated by the outer regions.

In the halo component, fluxes from the IRAS maps may be affected in an 
unknown way by instrumental spurious effects, as seen above. 
As explained  in www.ipac.caltech.edu/ipac/iras/iras\_data\_features.html, 
very bright ($>$ 500 Jy) pointlike sources may produce a six-pointed star shape 
due to the reflection from the secondary mirror struts; there is no known 
method to remove this effect.
Such a pattern is seen in the IRAS  maps of NGC 253 at 60\,$\mu$m and 100\,$\mu$m
(Fig.~\ref{fig:dispcont});  
however, we note that the same pattern in the `putative' diffuse emission 
may be recognized not only in the  ISOPHOT 180$\mu$m map, but also in the 
ROSAT images (see e.g.: Pietsch et al. \cite{pietsch}, panels B
(0.1-2.4 keV) and S (0.1-0.4 keV) in their Fig.1; Forbes et al. \cite{forbes}, 
Fig.2). 
We therefore conclude that the diffuse emission detected in the IRAS maps is 
probably real at $\lambda \ge 60\,\mu$m and maybe at 25\,$\mu$m, while no 
significant contribution is detected at 12\,$\mu$m; since we cannot  exclude
the existence of instrumental effects, fluxes from low surface brightness regions
in the IRAS maps should be considered as upper limits.

\begin{table*}
\caption{
Distances from nucleus (r) and diameters (d) of the apertures within
 which fluxes were measured;  dust
temperatures, infrared  luminosities (1-1000\,$\mu$m) and dust masses
derived from the fits with and without the 25\,$\mu$m point  are 
displayed.}
\label{tab:params}
\begin{flushleft} 
 \begin{tabular}{l @{\extracolsep{.1cm}}cccccc}
 \hline
\noalign{\smallskip}
& Total & Nucleus & \multicolumn{2}{c}{Disk} & \multicolumn{2}{c}{Halo} \\
\cline{2-2} \cline{3-3} \cline{4-5} \cline{6-7}
           &       a       &       b       &       c       &       d       &       e       &       f    \\ 
r [\arcmin] &	0	&	0	&	8	&	8	&	9	&	8	\\
r [kpc]       &  0	&	0	&	6	&	6	&	7	&	6	\\
d [\arcmin] &       33      &       8       &       5       &       5       &       8       &       5    \\ 
d [kpc]	&	24	&	6	&	4	&	4	&	6	&	4	\\

\noalign{\smallskip}
\multicolumn{3}{l}{Observed fluxes, not color corrected}\\
12\,$\mu$m   & 63.2   &     35.0   &     2.0  &   1.7   &  --   &  --  \\
25\,$\mu$m   & 181   &     178    &     2.4  &   1.5   &  3.3  &  $\le$ 0.1  \\
60\,$\mu$m   & 1186    &     853    &     19.1 &   12.2  &  6.1  &  3.6  \\ 
100\,$\mu$m  & 1770    &     949    &     56.6 &   44.0  &  25.5 &   8.3  \\ 
180\,$\mu$m  & 1693    &     607    &     87.3 &   77.4  &  28.3 &    15.7\\ 

\noalign{\smallskip}
\multicolumn{5}{l}{\emph{Fit from 25\,$\mu$m to 180\,$\mu$m}}\\

\multicolumn{5}{l}{Warm component}\\ 
T [K]      &       60      &       60      &       60      &       60      &       80      &       40   \\ 
$\mathrm{L_{ir}}$  $[\mathrm{L_{\sun}}]$  &    1.07E+10   &    8.18E+09   &    1.44E+08   &    9.08E+07   &    8.32E+07   &    3.33E+07\\
$\mathrm{M_d}$ $[\mathrm{M_{\sun}}]$          &    7.42E+04   &    3.59E+04   &    1.04E+03   &    6.47E+02   &    6.75E+01   &    1.61E+03\\
\multicolumn{5}{l}{Cold component}\\ 
T [K]      &       20      &       20      &       20      &       20      &       20      &       15   \\ 
$\mathrm{L_{ir}}$  $[\mathrm{L_{\sun}}]$  &    8.51E+09   &    4.24E+09   &    3.76E+08   &    3.14E+08   &    1.56E+08   &    5.80E+07\\
$\mathrm{M_d}$ $[\mathrm{M_{\sun}}]$         &    3.51E+07   &    7.23E+06   &    2.69E+06   &    2.79E+06   &    5.42E+05   &    9.84E+05\\

\noalign{\smallskip}
\multicolumn{5}{l}{\emph{Fit from 60\,$\mu$m to 180\,$\mu$m}}\\
			
\multicolumn{5}{l}{Warm component}\\													
T [K] & 47 & 39	& 28 &	27 & 25	& 29 \\
$\mathrm{L_{ir}}$  $[\mathrm{L_{\sun}}]$  & 7.90E+09 & 5.76E+09 & 1.91E+08 & 
1.27E+08 & 1.26E+08 & 4.60E+07 \\
$\mathrm{M_d}$ $[\mathrm{M_{\sun}}]$ & 2.36E+05  & 5.47E+05 & 1.34E+05 &
 1.05E+05 & 1.86E+05 & 2.32E+04 \\

\multicolumn{5}{l}{Cold component}\\													
T [K] & 20 & 20	& 17 & 16 & 17 & 13 \\
$\mathrm{L_{ir}}$  $[\mathrm{L_{\sun}}]$ & 7.72E+09 & 2.22E+09 & 2.51E+08 
& 2.29E+08 & 4.40E+07 &	 4.22E+07 \\
$\mathrm{M_d}$ $[\mathrm{M_{\sun}}]$  & 3.83E+07	& 9.81E+06 & 3.88E+06 &	 
3.85E+06 & 6.74E+05 & 3.30E+06	\\

\hline
 \end{tabular}
\end{flushleft}
\end{table*}

\subsection{Spectral Energy Distributions}
Physical quantities (dust temperatures and masses) 
in the different regions were first derived fitting the observed SED with 
modified blackbodies, 
i.e. blackbodies with a wavelength dependent emissivity
(see e.g. Hildebrand \cite{hildebrand}).
Far--infrared fluxes in the observed range (12-180\,$\mu$m) are probably 
due (Desert et al. \cite{desert}) to a mix of polyaromatic hydrocarbon 
molecules (PAHs), very small grains and big grains. 
Emission is dominated by PAHs at 12\,$\mu$m, by cold silicate and graphite 
grains at $\lambda \ge$ 140\,$\mu$m (Dwek et al. \cite{dwek}, Sturm et al. 
\cite{sturm}, Genzel \& Cesarsky \cite{genzel}); at intermediate wavelengths 
each of the three components may contribute to the observed fluxes.
Our purpose is to determine temperatures and masses of the cold dust 
component, but due to the limited number of points it is not possible to 
obtain a univoque fit (see Lisenfeld et al. \cite{lisenfeld} for a discussion 
about the non uniqueness of single-temperature fits for the far-infrared 
SEDs). 
In order to estimate the impact of this uncertainty we first made a fit 
including all the points with the exception of the one at 12\,$\mu$m, which is 
dominated by PAHs. The fit was then repeated dropping the 25\,$\mu$m point,
leading to lower dust temperatures and higher dust masses; values from this
fit are indicated in italics.

We used for the fit the sum of two  modified blackbodies and an emissivity 
$\propto \lambda^{-2}$. 
Color correction was applied according to the temperatures found and the whole 
procedure was repeated until convergence. The observed fluxes and fitted SEDs 
are shown in  Table~\ref{tab:params} and Fig.~\ref{fig:seds}. 
For each component, we computed 
dust masses according to Klaas \& Els\"asser (\cite{klaas}), assuming dust 
grain properties as in Hildebrand (\cite{hildebrand}) for the emissivity  
adopted here : 
\begin {equation}
M_{\rm d} = 7.9 \times  10^{-5} (T_\mathrm{K}/40)^{-6} \ L_{\rm IR}/L_\odot\ \ \ [M_\odot],   
\label{eq:hildebrand}
\end{equation}
where $L_{\rm IR}$ is the luminosity between 1 and 1000\,$\mu$m computed from the  
extrapolation of the modified black--bodies. 

Total fluxes from the whole galaxy ({\em a}) were computed from the 
integration in a 
circular aperture with 33\arcmin\ diameter. Fluxes from 25 to 180\,$\mu$m 
are well fitted by a two-component blackbody with T $\sim$ 60({\em 47}) K and 
T $\sim$ 20 K. 
Using the dust mass derived from the SED integrated over the whole galaxy
($M_{\rm d} \sim$ 3.5({\em 3.8})$\times10^7$ M$_\odot$) and the total mass 
($M({\rm HI+H_2}) = 2.4\times10^9$ M$_\odot$) estimated by Houghton et 
al. (\cite{houghton}) we  obtain a dust to gas mass ratio $\sim$ 1/70, 
somewhat higher than the Galactic value (1/160). 

\begin{description}
\item{{\em 1. Nucleus} --} The nuclear SED ({\em b}) is fitted by two components, 
a warm one (T $\sim$ 60({\em 39}) K) and a cold one (T $\sim$ 20 K). 
The ratio of the extrapolated 
1-1000 $\mu$m luminosities in the two components is $L_{w}/L_{c} \sim 2$, 
i.e. the emission is dominated by the warm component. We estimate a dust mass 
$\sim$ 7({\em 10})$\times10^6$ M$_\odot$. 
In order to check the reliability of the separation of the `nuclear' emission 
from the extended emission, we extrapolated the nuclear SED to 1300\,$\mu$m and 
compared it with the flux given by Kr\"ugel et al. (\cite{krugel}), 1.7 Jy in 
an area of $3\arcmin\times2\arcmin$: the two values agree within 30({\em 8})\%, 
which is acceptable considering the uncertainties involved. 
\item{{\em 2. Disk} --}  
The SEDs in the NE ({\em c}) and SW ({\em d}) regions of the disk show, as in 
the nucleus, the presence of a warm (T $\sim$ 60({\em 28}) K) and a cold 
(T $\sim$ 20({\em 17}) K) component. 
However, emission is dominated by the cold component, 
$L_{w}/L_{c} \sim$ 0.3({\em 0.5}). 
The dust mass estimated in each of these regions is 
$\sim$ 3({\em 4})$\times10^6$ M$_\odot$.
\item{{\em 3. Halo} --}  Due to the uncertainty in the IRAS fluxes, a SED
decomposition is very difficult; the results given here are therefore indicative.
In the southern halo ({\em f}) we could only assume an upper 
limit for the flux at 25\,$\mu$m since no significant emission was detected.
Dust temperatures of the cold component are similar to those found in the disk, 
$T \sim$ 20({\em 17}) K, with a somewhat lower temperature in  the southern halo
 ({\em f}) where $T \sim$ 15({\em 13}) K; we derive dust masses in the order of  
0.5-1({\em 0.7-3})$\times10^6$ M$_\odot$.
In order to explain the non
detection by IRAS of emission by dust involved in outflows, Alton et al. 
(\cite{alton-farir}a) argued that the temperature of the dust is  $T \sim 
15-20$ K and its mass  $< (0.5-3.5)\times10^5$ M$_\odot$: this limit is not far
from the values we find for \object{NGC 253}, in particular if we consider that if 
$T \sim 20$K an uncertainty in the temperature of $\Delta T=5$K may change the 
dust mass by a factor of 5.  
For comparison, Phillips (\cite{phillips}) gave a dust mass of 
$2\times10^5$ M$_\odot$ for the 
outflow in \object{NGC 1808}; Alton et al. (\cite{alton-submm}) estimated for 
the outflow in M82 dust masses of $10^6-10^7$ M$_\odot$.   
\end{description} 
 
According to Kennicutt (\cite{kenn}), the star formation rate for continuous 
bursts of age 10-100 Myr in starbursts, where 
contribution of dust heating from old stars may be neglected, is:
\begin{equation}
{\rm SFR} = 1.7 \times 10^{-10} L_{\rm IR}/L_\odot\ \ [M_\odot\ \rm{yr}^{-1}].
\end{equation}
The SFR obtained in the nucleus and
in the whole galaxy is $\sim$ 2.1({\em 1.3}) and 
3.3({\em 2.6}) M$_\odot\ \rm{yr}^{-1}$, respectively. 
This low value of the SFR is in agreement with the conclusion by Engelbracht et 
al. (\cite{engelbracht}) that \object{NGC 253} is in a late phase of the 
starburst, having passed a rapid decrease of the star formation rate.  

\section{Modeling the FIR emission}
\label{sec-models}

The far--infrared emission of \object{NGC 253} was modeled with a
Monte Carlo simulation system (Kahanp\"a\"a 2001, in preparation)
using both the 180\,$\mu$m map for FIR major and minor axis
profiles and the strip maps at 0, 10 and 15\arcmin\ for minor axis 
profiles and photometry at 120 and 180\,$\mu$m.

\subsection{The galaxy model}
The galaxy model is based on a cylindrically symmetrical disk with exponential
distributions along the radius and height with an additional, compact 
nuclear energy source. The distribution of stars within
the disk was assumed to be independent of wavelength; no intrinsic color
gradients in the stellar population were taken into account. 
While this
ignores color gradients due to the bulge, the impact on the resulting FIR
colors is minor as \object{NGC 253} is a disk-dominated spiral (Baggett et al.
\cite{baggett}).  
The core was treated as an independent source with its own unobscured spectral
energy distribution.
Since recent studies by Davies et al. (\cite{davies99})
suggest that the scale  lengths of star and dust populations are not equal, the
scale lengths were treated  as independent parameters in the model. 

\subsection{Radiative transfer code} 
Unlike most radiative transfer codes our model does not ignore the cloud
structure of the interstellar medium (ISM). 
Instead, all dust is in clouds;  the total number of clouds is
determined by the total dust mass and the cloud positions are randomly
chosen from a double-exponential distribution:
\begin{equation}
R_1 = \frac{\int_0^r u \mathrm{e}^{-u/r_0}\mathrm{d}u}{\int_0^{r_{max}} u
\mathrm{e}^{-u/r_0}\mathrm{d}u}
,
\end{equation}
\begin{equation}
R_2 = \frac{\int_{-z_{max}}^z \mathrm{e}^{-u/z_0}\mathrm{d}u}{\int_{-z_{max}}^{z_{max}} 
\mathrm{e}^{-u/z_0}\mathrm{d}u}
, 
\end{equation}
where $\mathrm{R}_1$ and $\mathrm{R}_2$ are uniform random numbers
and $r_0$ and $z_0$ the dust scale length and scale height respectively.

The radiative transfer code for the optical domain is based on
models by de Jong (\cite{deJong}) and Bianchi et al. (\cite{bianchi}).
The assumed unobscured spectral energy distribution of the energy sources
(stars) was divided into
a number of energy bins -- each formally representing
one broad-band filter -- and the radiative transfer problem was numerically
solved for each bin by following numerous simulated photons through the dust
system. For each bin $2\cdot 10^5$ to $1\cdot 10^6$ random paths through the
dusty layer were calculated; the total number of independent paths used in a
single model was typically $40\cdot 10^6$. The resulting surface brightness
distribution for each wavelength was recorded. Additionally the cumulative
amount of absorbed energy at each position within the model galaxy was
registered for further processing. When all bins had been processed, the
equilibrium temperature of individual dust grains as a function of grain size
and position within the galaxy (distance from the nucleus and height above the
plane of the galaxy) was solved. In thermal equilibrium the emitted power from
a single grain is
\begin{equation}
P_{em} = \int_0^\infty 4 \pi^2 \mathrm{a^2\, Q_{abs}(\lambda,a)} \,
\mathrm{B(\lambda,T)}\;\mathrm{d}\lambda,
\end{equation}
where $\mathrm{a}$ in the grain radius, $\mathrm{Q_{abs}}$ is the absorption
coefficient of the grain and $\mathrm{B}$ is the blackbody (Planck) function.
The shape of $\mathrm{Q_{abs}}$ in the FIR was approximated with
\begin{equation}
\mathrm{Q_{abs}(a,\lambda)} =  1.8\times10^{-4}\mathrm{a}\,\lambda^{-\beta}.
\end{equation}

We have used $\beta = 2$, a typical value for grain temperatures less than 50
K, and dust grain size distributions and optical properties from  D\'esert
et al. (\cite{desert}). The dust model chosen is of secondary importance; we
tried using the classical MRN dust population models with $n(a)\propto
a^{-3.5}$ (Mathis et al. \cite{Mathis77}) and got the same results for
classical dust emission. Infrared emission from very small grains
and PAHs with stochastic heating and cooling processes were neglected as they
are minor contributors in the  100-200\,$\mu$m wavelength range studied here;
 they were, however, still included as absorbers with size distributions
and optical properties as given by D\'esert et al. (\cite{desert}).

\begin{figure}
      \resizebox{\hsize}{!}{\includegraphics{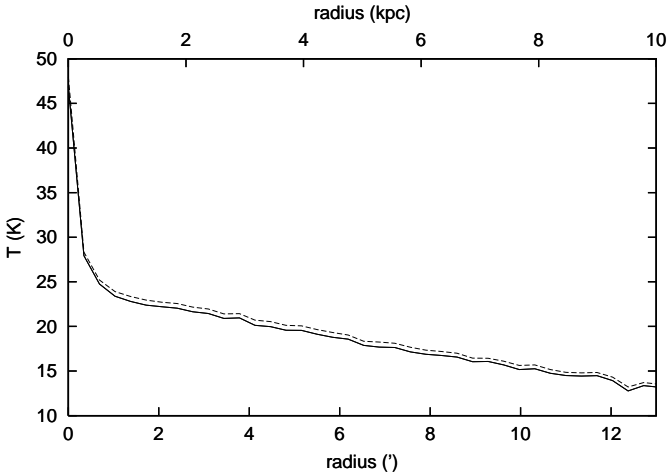}}

      \caption{Grain temperature in the galactic plane of \object{NGC 253}. 
	   The solid line represents grains with a diameter of 150 nm, the
	   dotted line the smallest classical grains in the model (d = 15
	   nm).}
      \label{FigT}
\end{figure}

\subsection{Model input parameters} 
Table~\ref{tab:model} summarizes the parameter values used in the model. The
observed total stellar fluxes as a function of  wavelength were obtained from
the NASA-IPAC extragalactic database (NED). The radial scale length of the
stellar component was obtained by comparing the resulting brightness profiles
for the optically thin outer parts of the axial profiles  with published
measurements of the major and minor axis profiles in the B and V bands (Pence
\cite{pence}, Baggett et al. \cite{baggett}). As a byproduct the inclination of
\object{NCG 253} was found to be 78$\degr \pm$ 1$\degr$. Limiting values for
the radial dust scale length were found by comparing the model results with the
observed major axis profile at 180\,$\mathrm{\mu m}$. Since the model profile
depends also on the temperature of the dust grains, this fit does not give as
stringent limits as  in  the optical case, but if the scale length was
varied by more than 15\%, the fit was unacceptably poor. The major parameters
of the model were adjusted until a reasonable fit to the optical and FIR data
was found. Unobscured SEDs for the disk and nucleus components were 
approximated with model SEDs for stellar populations (Silva \cite{silva}). 
For the core we used a 200-My old population without modifications; in the disk
a combination of young (0.2 Gy) and old (5 Gy)  populations was required. 
The resulting unobscured SED is shown in Fig.~\ref{FigTotSp}.

\begin{table}
\caption{Model input parameters and results for NGC 253}
\label{tab:model}
\begin{flushleft} 
 \begin{tabular}{l @{\extracolsep{1cm}}l}
 \hline
\noalign{\smallskip}
\multicolumn{2}{l}{\emph{Input parameters}}\\
\noalign{\smallskip}
  Distance            & 	2.5 Mpc \\       
\noalign{\smallskip}
  Stars, luminosity   & 	$3.2 \times 10^{10}\;\mathrm{L}_{\sun} $\\
  Stars, core to disk luminosity ratio  & 0.5 \\
\noalign{\smallskip}
  Dust, cloud size    & 	40 pc\\
  Dust, grain density & 	3 $\mathrm{g\, cm^{-3}}$\\
  Dust, maximum albedo & 	0.7 \\
  Dust, size distribution & As in D\'esert et al. \cite{desert} \\
\noalign{\smallskip}
\multicolumn{2}{l}{\emph{Results}}\\
  Inclination	      & 	78\degr \\
\noalign{\smallskip}  
  Stars, scale length & 	1.8 kpc \\
  Stars, scale height & 	0.2 kpc \\
\noalign{\smallskip}
  Dust, scale length  & 	2.8 kpc\\
  Dust, scale height  & 	0.15 kpc\\
  Dust, total mass    & 	$2.8 \times 10^7 M_{\sun}$ \\
  $\mathrm{\tau _V(center)}$ &  4.5 $\pm$ 0.5 \\
  FIR luminosity      & 	$1.7 \times 10^{10}\;\mathrm{L}_{\sun} $\\
  Large grain Lumin.  & 	$1.2 \times 10^{10}\;\mathrm{L}_{\sun} $\\
  VSG \& PAH Lumin.   & 	$0.45 \times 10^{10}\;\mathrm{L}_{\sun} $\\
  T(r=0.5 kpc)	      & 	25 $\pm$ 1 K  \\
  dT/dr in disk       & 	1.2 K\,kpc$^{-1}$\ \ (1­-10 kpc)\\
\hline
\end{tabular}
\end{flushleft}
\end{table} 
%
%
\subsection{Predicted dust mass, distribution and temperature profiles}
\label{subsec-mdt}
According to the computed model the face-on optical depth of the dust layer at
the center of the galaxy is 4.5 at 550 nm. After the inclination 
of \object{NGC 253} (78\degr ) is taken into account  
this makes the core attenuated by roughly 11 magnitudes in the V band.
Values based on earlier observations are somewhat higher: Engelbracht et al.
(\cite{engelbracht}) estimated \mbox{$\mathrm{A_V(cen)}$ = 19.07 $\pm$ 3.3}
from the H-K excess, but as the small starburst core certainly includes a 
sizeable amount of hot local and circumstellar dust, this is to be
expected. Half (54\%) of the total model luminosity \mbox{($3.2 \times
10^{10}\, \mathrm{L}_{\sun}$)} is converted to FIR emission by the 
dust grains. The total mass of the cold dust component as inferred from the
model is \mbox{$\mathrm{M}\sim 0.93\times 10^7 \rho_\mathrm{d}\; M_{\sun}$}, where
$\rho_\mathrm{d}$ is the average density of the dust grains in units of
\mbox{$\mathrm{g\,cm^{-3}}$}. Using the value ($\rho_\mathrm{d} = 3\;\mathrm{
g\,cm^{-3}}$) given by Hildebrand (\cite{hildebrand}) sets the dust mass of the
model to \mbox{$2.8 \times 10^7 M_{\sun}$}, a value not too far from the
estimate found by using Eq.~\ref{eq:hildebrand}.

A disk-to-nucleus luminosity ratio of 2.0 gave the best match with the observed SED
and  also results in a core luminosity $\mathrm{L_{core}} = 1.1 
\times 10^{10} \mathrm{L}_{\sun}$; 
the same value was found by Engelbracht et al.
(\cite{engelbracht}) from a combination of mid- and far-infrared,
sub-millimeter and 1.3 mm data.

Fig.~\ref{FigT} shows the radial temperature distribution of dust in the 
plane of \object{NGC 253}. The smallest grains tend to be hotter  and thus more
efficient FIR emitters than their bigger companions because the FIR absorption
coefficient of the grains at a given wavelength is proportional to grain
diameter (see Draine \& Lee \cite{draine}). This effect is partially
counterbalanced by the spectral shape of the attenuated interstellar radiation
field (ISRF) within the galaxy: the smallest grains are poor absorbers in the
near infrared, where the ISRF is most intense.

\begin{figure}
      \resizebox{\hsize}{!}{\includegraphics{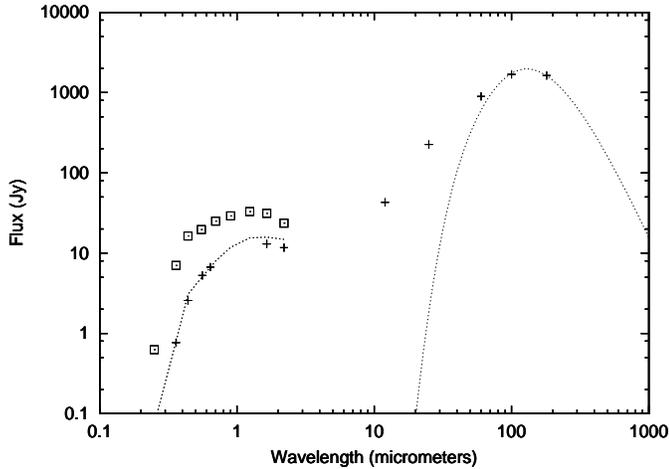}}
      \caption{Spectral energy distribution of NGC 253 in 
	the optical-near infrared (ground-based photometry from  NED) and
	far--infrared (IRAS 100\,$\mu$m and ISOPHOT 180\,$\mu$m fluxes).  
	Crosses -- observed SED; squares -- unobscured SED;  dashed line
	--  model fit for the emission from stars and cold dust, as described
	in Sect.~\ref{sec-models}. IRAS fluxes from 12 to 60\,$\mu$m were
	ignored in the fit since the model does not include the IR
	emission from PAHs or very small  grains, which are responsible for
	most of the emission at these  wavelengths.
	}
      \label{FigTotSp}
\end{figure}

The central dust temperature given by the model of \object{NGC 253} is 
higher than that given by a simple fit to the FIR spectrum 
(Fig.~\ref{fig:seds}). The
central spectrum is, however, integrated over quite a large area and in the end
the fit is sufficiently good.  The hot central core 
is very small and its contribution to the total SED is approximately 40\% of the 
total luminosity. The total SED (Fig.~\ref{FigTotSp})  as well as 
the strip map flux levels at 120 and 180\,$\mu$m at 0\arcmin\ and 10\arcmin\ 
are perfectly reproduced.

The derived scale lengths of the stellar and dust components are 1.8 and 2.8
kiloparsecs respectively. As the uncertainty of both values is of the order of
15\%, the difference is significant. This agrees with the recent observations
of dust distribution in other nearby spiral galaxies (Alton et al.
\cite{alton98}b, Alton et al. \cite{alton01}).
A prominent effect of the difference in the scale lengths is the variation of
the dust to stellar mass ratio with radius: at 10 kpc the relative dust density
is four times as large as at the center.

The halo dust component cannot be reproduced by a single 3-D exponential dust
disk: the modeled minor axis profiles (convolved to 120\arcsec\ FWHM) are 
narrower than the observed profiles of \object{NCG 253} at 120 and 180\,$\mu$m.

This can be due to several different effects or a combination of them: 
\begin{enumerate} 
  \item There is an extra source heating the halo, which is not 
     represented by the stellar populations in the model.
  \item A halo population of young stellar objects (YSO) or other 
     compact FIR emitters is present in NGC 253. 
  \item The dust distribution in the halo has a distribution which 
     significantly deviates from a simple exponential model. 
\end{enumerate} 

The two first scenarios are in agreement with the existence of star  formation
in the halo of NGC 253, which has been recently proposed by Comer\'on et al. 
(\cite{comeron}) on the basis of UBV photometry. In this case we
should be able to see the optical  emission of the heating sources as the halo
is certainly optically thin. The very cold spectrum of the outer parts
contradicts the second case as YSO sources usually have SED with 
strong near and mid-infrared components.
In the third case, which we find  most probable,
the excess dust in the halo is heated by stars in the disk of NCG 253. If 
a dust halo is the most important contributor it should also be observable 
in the optical domain as a faint, blue halo of scattered light.

\section{Conclusions}

ISOPHOT observations  were used together with IRAS data to map the far-infrared
emission in  the starburst galaxy \object{NGC 253}. The resulting images were
then analyzed with modified blackbody fits as well as radiative transfer
modeling. The total FIR luminosity consists of two parts: a warm component (T
$\sim$ 47-60 K), dominated by the central regions, and a cold component 
(T $\sim$ 20 K)
produced in the disk and the halo. Dust temperatures and masses derived from
single-temperature fits of the SED integrated over the whole galaxy are
in agreement with those obtained by solving the radiative transfer equations.
The results obtained in the different regions may be summarized as follows:

{\em Nucleus}: the nucleus produces half of the total FIR luminosity of NGC
253. Our fit for the core correctly predicts the observed 1.3 mm continuum
emission given by Kr\"ugel et al. (1990). The low inferred star formation rates
suggest that the nuclear starburst in \object{NGC 253} is in a late phase,
having passed a rapid decrease of the  star formation rate.

{\em Disk}: the resulting dust temperatures are typical for the general diffuse
ISM in spiral galaxies. The radiative transfer model indicates that the dust
scale length is $\sim$ 40\% larger than that of stars; the same trend was found
by Alton et  al. (\cite{alton98}b). The dust is warmer ($T > 20$K) in the
central regions of  the galaxy, the outer regions are dominated by colder dust
($T \le 15$K).

{\em Halo}: we find evidence for extended emission to projected distances of 
$\sim$ 10 kpc from the nucleus along the minor axis both in the IRAS and
ISOPHOT  maps; the emission seen in the IRAS maps could be due to instrumental
effects, but since the same emission pattern is also seen in the ISOPHOT map 
as well as in ROSAT images (e.g. Pietsch et al. \cite{pietsch}, Forbes et al. 
\cite{forbes}) we conclude that this is not the case. The SED fitting of this 
extended component is very uncertain, since IRAS fluxes {\em may still be
affected  by instrumental effects}. Keeping in mind this uncertainty, the
emission is  probably dominated by cold dust ($T \sim 15-20$ K); the derived
dust masses are  comparable to those found for the outflows in NGC 1808 and M
82. The apparently constant mass in the outflows observed in  nearby galaxies
is probably due to limitations in the instrumental resolution and  sensitivity.
It was not possible to reproduce the halo component by a single 3-D
exponential  dust disk.  Even if in our opinion this is most likely due 
to a non exponential distribution of dust in the halo, some extra heating may 
be given by star formation, which according to Comer\`on et al. 
(\cite{comeron}) may occur in the halo. Both phenomena may be related to the 
interaction of the interstellar medium with the superwind driven by the 
nuclear starburst.

\begin{acknowledgements}
The ISOPHOT development and the postoperation phase
performed in the ISOPHOT Data Centre at the  Max-Planck-Institut f\"ur Astronomie 
Heidelberg are supported by Deutsches  Zentrum f\"ur Luft- und Raumfahrt  (DLR), 
Bonn. The authors are responsible for the content of this paper.
PIA has been jointly developed by the ESA Astrophysics Division and the ISOPHOT 
consortium. We are grateful to the referee for the useful comments which 
improved this paper. We are indebted to
Manfred Stickel of the Heidelberg ISOPHOT Data Center for having provided his
procedures. We thank IPAC for the IRAS HIRES maps. This research has made use
of the NASA-IPAC extragalactic database (NED) which is operated by the Jet 
Propulsion Laboratory, Caltech, under contract with the NASA. 
\end{acknowledgements}


\end{document}